\documentclass[%
 reprint,
 amsmath,amssymb,
 twocolumn,
prl,
floatfix,
superscriptaddress,
]{revtex4-2}

\usepackage[english]{babel}
\usepackage{graphicx}
\usepackage{dcolumn}
\usepackage[utf8]{inputenc} 
\usepackage{bm}
\usepackage[colorlinks=true,citecolor=blue,linkcolor=blue,urlcolor=blue]{hyperref}
\usepackage{mathrsfs}
\usepackage{longtable}
\usepackage{multirow,booktabs,dcolumn,tabularx}
\usepackage{float}

\usepackage{braket}
\usepackage[dvipsnames]{xcolor}

\usepackage{amsmath,amsfonts,amsthm} 


\begin{document}

\newcommand{\ptpb}{PtPb$_4$ }
\newcommand{\pnbm}{\textit{P}4/\textit{nbm}}

\title{Superconducting density of states of PtPb$_4$}

\author{Pablo Garc\'ia Talavera}
\affiliation{Laboratorio de Bajas Temperaturas y Altos Campos Magn\'eticos, Departamento de F\'isica de la Materia Condensada, Instituto Nicol\'as Cabrera and Condensed Matter Physics Center (IFIMAC), Unidad Asociada UAM-CSIC, Universidad Aut\'onoma de Madrid, E-28049 Madrid, Spain}

\author{Jose Antonio Moreno}
\affiliation{Laboratorio de Bajas Temperaturas y Altos Campos Magn\'eticos, Departamento de F\'isica de la Materia Condensada, Instituto Nicol\'as Cabrera and Condensed Matter Physics Center (IFIMAC), Unidad Asociada UAM-CSIC, Universidad Aut\'onoma de Madrid, E-28049 Madrid, Spain}

\author{Edwin Herrera}
\affiliation{Laboratorio de Bajas Temperaturas y Altos Campos Magn\'eticos, Departamento de F\'isica de la Materia Condensada, Instituto Nicol\'as Cabrera and Condensed Matter Physics Center (IFIMAC), Unidad Asociada UAM-CSIC, Universidad Aut\'onoma de Madrid, E-28049 Madrid, Spain}

\author{Alexander I. Buzdin}
\affiliation{LOMA UMR-CNRS 5798, University of Bordeaux, F-33405 Talence, France}
\affiliation{Institute for Computer Science and Mathematical Modeling, Sechenov First Moscow State Medical University, 119991 Moscow, Russia}

	\author{Sergey L. Bud'ko}
	\affiliation{Ames Laboratory and Department of Physics \& Astronomy, Iowa State University, Ames, IA 50011}
		
	\author{Paul C. Canfield}
	\affiliation{Ames Laboratory and Department of Physics \& Astronomy, Iowa State University, Ames, IA 50011}

\author{Isabel Guillam\'on}
\affiliation{Laboratorio de Bajas Temperaturas y Altos Campos Magn\'eticos, Departamento de F\'isica de la Materia Condensada, Instituto Nicol\'as Cabrera and Condensed Matter Physics Center (IFIMAC), Unidad Asociada UAM-CSIC, Universidad Aut\'onoma de Madrid, E-28049 Madrid, Spain}

\author{Hermann Suderow}
\affiliation{Laboratorio de Bajas Temperaturas y Altos Campos Magn\'eticos, Departamento de F\'isica de la Materia Condensada, Instituto Nicol\'as Cabrera and Condensed Matter Physics Center (IFIMAC), Unidad Asociada UAM-CSIC, Universidad Aut\'onoma de Madrid, E-28049 Madrid, Spain}

\begin{abstract}
PtPb$_4$ is a type II superconductor with a bulk critical temperature $T_{c}\approx 3 $\ K and an upper critical field of $H_{c2}=0.36 $\ T. PtPb$_4$ is related to non-superconducting PtSn$_4$, which presents nodal arc states at the surface. Here we measure the superconducting density of states of PtPb$_4$ using millikelvin Scanning Tunneling Microscopy (STM). We observe a fully opened superconducting gap of $\Delta=0.48$\ meV similar to expectations from Bardeen Cooper and Schrieffer (BCS) theory ($\Delta_0=1.76k_BT_{c}=0.49 $\ meV). Measurements under magnetic
fields applied perpendicular to the surface show a spatially inhomogeneous gap structure, presenting superconducting signatures at fields as high as 1.5 T, significantly above $H_{c2}=0.36 $\ T. On some locations we find that the superconducting density of states does not vanish above $T_{c}$. We can find signatures of a superconducting gap up to 5 K. We discuss possible reasons for the observation of superconducting properties above $T_{c}$ and $H_{c2}$, emphasizing the role played by structural defects.
\end{abstract}

\maketitle

\section*{Introduction}

Recent interest in topological semimetals is driven by the discovery of non-trivial surface states presenting Dirac or Weyl dispersion when surface bands touch at points or closed contours\,\cite{RevModPhys.82.3045,10.1093/nsr/nwae272}. These materials often present an extremely large magnetoresistance\,\cite{Ali2014,Mun2012,PhysRevResearch.2.022042,doi:10.1073/pnas.1808747115,Liang2015,Shekhar2015,Wu2016}. The observation of a resistance enhancement by $\approx 5 \times 10^5\%$ with no indications of saturation up to 14 T in PtSn$_4$\,\cite{Mun2012}, led to the identification of novel Dirac nodal arcs in this compound\,\cite{Wu2016}. The Dirac nodal arc is distinct from the point and closed contour touchings found in several topological semimetals\,\cite{annurev:/content/journals/10.1146/annurev-conmatphys-031016-025458,RevModPhys.90.015001}. This discovery has stimulated the search for materials that simultaneously host Dirac nodal arcs and superconductivity, as such systems could exhibit exotic surface phenomena associated with topological superconductivity\,\cite{RevModPhys.83.1057,Alicea_2012,doi:10.1021/acs.chemmater.3c00713,Sato_2017}.

Among the compounds related to PtSn$_4$, there is superconducting AuSn$_4$ (with a critical temperature $T_c = 2.35 K$). AuSn$_4$ shows peculiar behavior at the surface, characterized by and anomalous low energy density of states within the gap and a critical temperature $T_c$ enhanced by about 20\% from the bulk\,\cite{Herrera2023,Zhu2023,Sharma2022,Shen2020}. Structural analyses have revealed the presence of stacking faults, which induce localized structural distortions and may be associated with nearly two-dimensional superconductivity\cite{Herrera2023,Shen2020}. Rashba spin-split bands, suggesting topological superconductivity, have been also observed\cite{Zhu2023}.

Similarly, the compound \ptpb is also superconducting, with $T_c \approx 2.8$\,K\,\cite{Gendron1962,Wang_2021,Xu2021,Lin2021}. \ptpb was originally reported to crystallize in a tetragonal structure which consists of layers formed by Pb-Pt-Pb groups, similar to the Sn-Pt-Sn layers of PtSn$_4$\,\cite{RoslerSchubert_1951,Long2009}. Within the tetragonal structure, nearly flat degenerate bands have been observed close to the Brillouin zone boundary arising from nonsymmorphic symmetry elements\,\cite{Wu_2022}. Nevertheless, recent work in \ptpb shows evidence for crystallization in the same orthorhombic structure of PtSn$_4$, and exhibits a twofold symmetry in its surface band dispersion\,\cite{Lee2021}. Furthermore, there is significant Rashba spin splitting at the surface\,\cite{Lee2021}. Measurements in the superconducting phase including magnetization and upper critical field studies, indicate that PtPb$_4$ is a type II superconductor with $H_{c2}\approx 0.36$ T and a well defined specific heat jump near $T_c$\,\cite{Xu2021,ShenPhD2020}. However, specific heat measurements at temperatures well below $T_c$ are lacking, leaving the low energy electronic properties of the superconducting phase largely unexplored. Given these findings, a detailed investigation of the superconducting properties at the surface of PtPb$_4$ is essential.

Here we provide Scanning Tunneling Microscopy (STM) measurements of the superconducting density of states down to 0.1 K, significantly below T$_c$. We find a fully open s-wave BCS superconducting density of states. There are signatures of a superconducting density of states at temperatures and magnetic fields exceeding the bulk $T_c$ and $H_{c2}$.

\section*{Experimental}

PtPb$_4$ crystals were grown following Ref.\,\cite{Lee2021}, using Pb flux\cite{CanfieldCrucible2016, CanfieldCrucible, Canfield01061992}. To perform the STM measurements we glued a single crystal onto a sample holder, mounted it into the STM and installed the whole system in a dilution refrigerator of Oxford Instruments. We cleaved the sample in-situ, below liquid helium temperature, by moving the sample holder below a beam, as described in Refs.\,\cite{10.1063/1.3567008,10.1063/5.0059394}. We describe the STM control, data acquisition software and further details of the set-up in Refs.\,\cite{10.1063/5.0064511,MONTOYA2019e00058}. We also use our image rendering software available in Ref.\,\cite{githublbtuam} and usual STM image rendering software\,\cite{10.1063/1.2432410,githublbtuam}. The samples were plate-like with an optically flat surface. The most likely cleaving plane is between the Pb-Pt-Pb blocks and should lead to an exposed Pb surface.  The magnetic field was applied perpendicular to the surface. We did not obtain atomic resolution, nor were able to image the vortex lattice, presumably due to the frequent appearance of rough surfaces with grains several tens of nm in size.

\begin{figure}[h]
    \centering
    \includegraphics[width=0.7\linewidth]{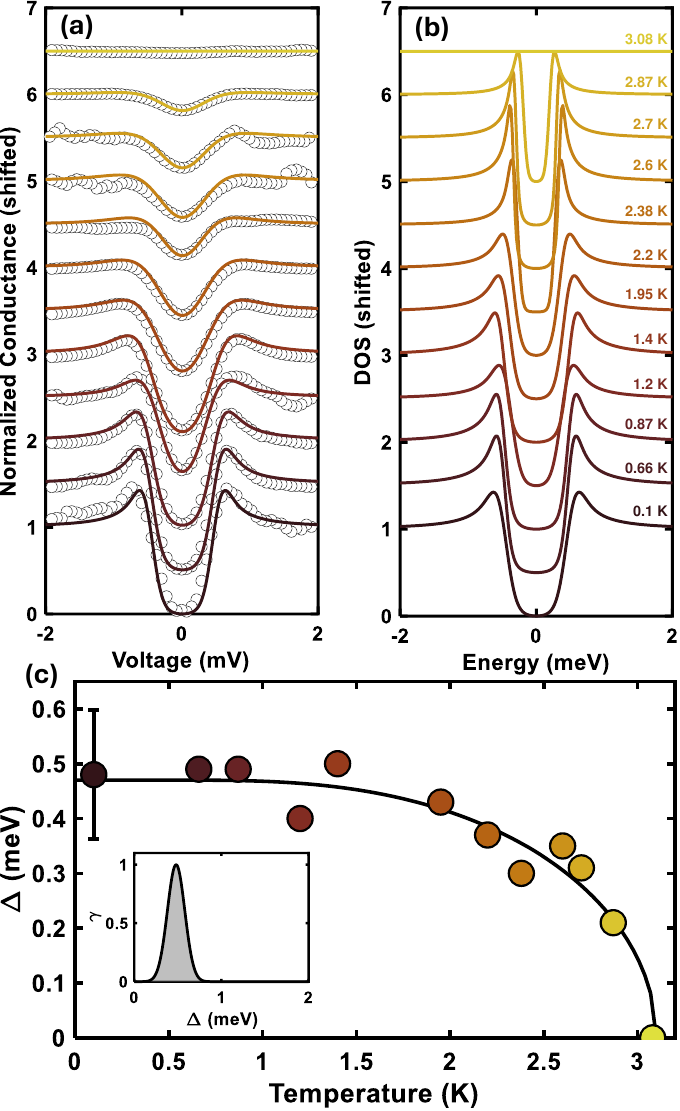}
    \caption{(a) We show as colored lines the tunneling conductance versus bias voltage curves as a function of temperature. The black lines are the fits obtained by convoluting the density of states curves shown in (b) with the derivative of the Fermi function. The temperature is shown close to each curve. (b) Density of states obtained as described in the text. The distribution of values of the superconducting gap $\gamma_i$ is shown as an inset in (c). (c) Superconducting gap $\Delta$ as a function of temperature. The value of $\Delta$ is determined by the position of the maximum in the distribution $\gamma_i$. The error bar represents the distribution of values of the superconducting gap, which remains roughly constant as a function of temperature. The black line represents the expected BCS tendency for a superconducting gap $\Delta_0=0.48$ meV.}
    \label{fig:GapvsT1}
\end{figure}

\section*{Results}

\begin{figure}[h]
    \centering
    \includegraphics[width=1\linewidth]{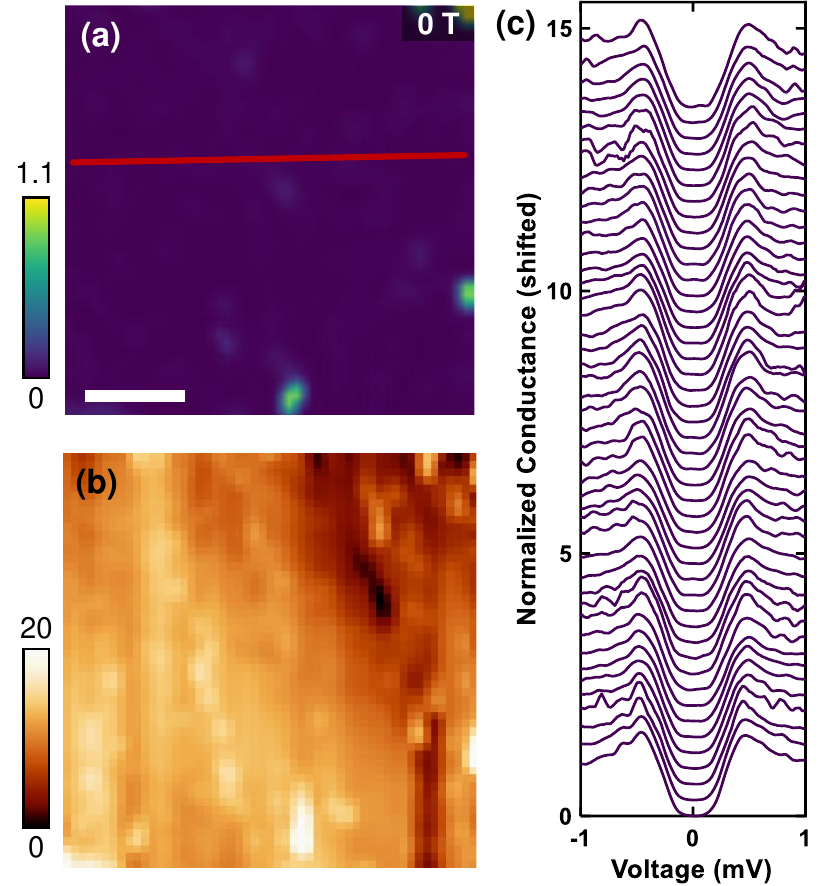}
    \caption{(a) Conductance map at zero bias and 0 T, at a temperature of 0.10 K. White scale bar is 50 nm large. Scale bar at the left is in zero bias conductance normalized to the conductance at voltages well above the gap. (b) Topography in the same field of view. Color scale on the left provides the height changes in nm. (c) Conductance curves taken along the red line in (a). Curves are colored by their zero bias value with the color scale of (a).}
    \label{fig:NoFieldCascade}
\end{figure}

In Fig.~\ref{fig:GapvsT1}(a) we show the superconducting density of states found over large areas on the cleaved PtPb$_4$ surface. We see that the gap is fully open, and there is a negligible density of states close to the Fermi level, as well as somewhat rounded quasiparticle peaks. Following the observation by angular resolved photoemission of several electron and hole bands crossing the Fermi surface\,\cite{Lee2021}, we assume a distribution of values for the superconducting gap $\Delta_i$ to obtain the superconducting density of states $N(E) \propto \sum_i \gamma_i(\Delta_i) \text{Re}\Big(\frac{E}{\sqrt{E^2-\Delta_i^2}}\Big)$. A similar procedure was succesfully used previously in other superconductors with multiple bands crossing the Fermi level\,\cite{10.1063/5.0059394,PhysRevB.92.054507,PhysRevResearch.4.023241,Herrera2023,PhysRevLett.96.027003,PhysRevLett.101.166407,PhysRevB.97.134501}. $\gamma_i(\Delta_i)$ is a Gaussian shaped distribution of the values of the superconduting gap. We then convolute $N(E)$ with the Fermi function at the measurement temperature (0.1 K) to obtain the tunneling conductance. To achieve the best agreement with experimental data, we test different forms of the distribution $\gamma_i(\Delta_i)$. The results are shown in Fig.~\ref{fig:GapvsT1}(b). We find that a Gaussian distribution centered at $\Delta_0=0.48$ meV with a width $\sigma_1=0.1$ meV provides the best fit (see inset of Fig.~\ref{fig:GapvsT1}(c)). The distribution remains similar for all temperatures. We can follow the maximum of the distribution with temperature and we obtain the temperature dependence of the superconducting gap $\Delta(T)$, (Fig.~\ref{fig:GapvsT1}c). $\Delta(T)$ follows BCS theory, with $T_{c}\approx 3$ K, close to the reported bulk value of $T_{c} \approx 2.8$\,K\,\cite{Gendron1962,Wang_2021,Xu2021,Lin2021}. $\Delta_0$ is also close to the expected BCS value $\Delta=1.76k_BT_{c}=0.49$ meV, with $T_{c}= 3$ K. The superconducting density of states is spatially homogeneous over large regions (Fig.~\ref{fig:NoFieldCascade}) and the distribution of values of the superconducting gap remains of the same order on the whole surface.

\begin{figure}[h]
    \centering
    \includegraphics[width=1\linewidth]{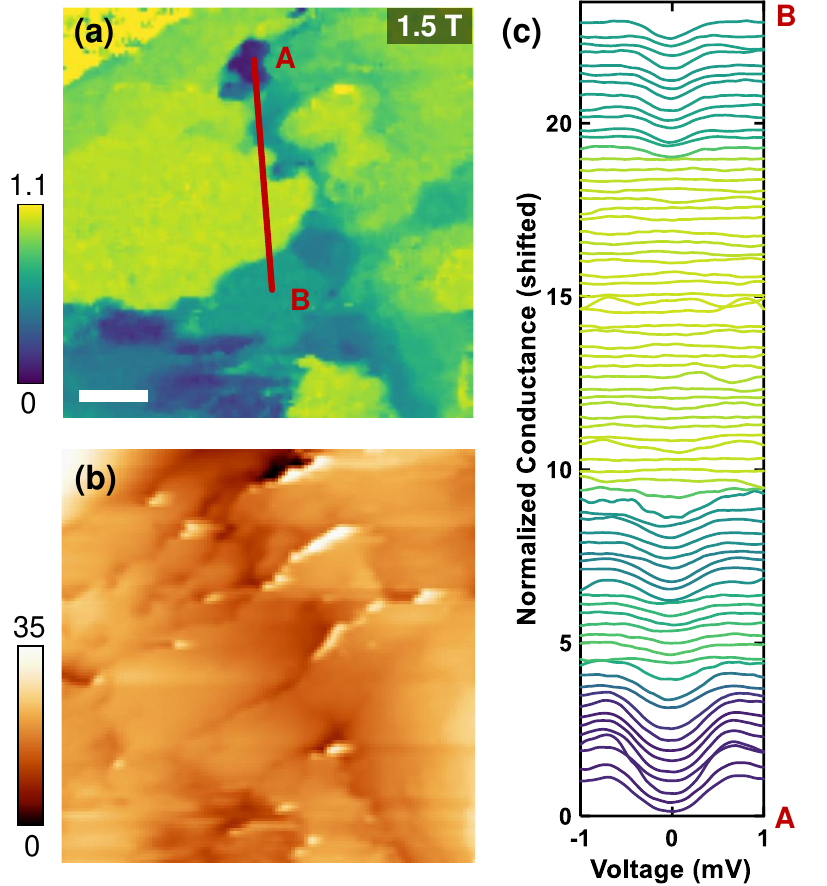}
    \caption{(a) Conductance map at zero bias and 1.5 T, at a temperature of 0.10 K. White scale bar is 50 nm large. Scale bar at the left is in zero bias conductance normalized to the conductance at voltages well above the gap. (b) Topography in the same field of view. Color scale on the left provides the height changes in nm in the same field of view. (c) Conductance curves taken along the red line in (a). We remark the position of top and bottom curves with A and B. Curves are colored by their zero bias value using the color scale of (a).}
    \label{fig:FieldCascade}
\end{figure}

When a magnetic field exceeding bulk $H_{c2}$ (Fig.~\ref{fig:FieldCascade}) is applied at 0.10 K, the surface predominantly exhibits a mixture of superconducting and normal regions. The normal regions display a flat tunneling conductance. The normal and superconducting areas are distributed in patch-like pattern across the surface. A correlation is evident between the spatial variations in the superconducting tunneling conductance at zero bias (Figs.~\ref{fig:FieldCascade}(a)) and the grain structure observed in the surface topography map (Figs.~\ref{fig:FieldCascade}(b)). It is remarkable that the superconducting gap can be well developed on top of some grains at such high magnetic fields (blue areas in Fig.~\ref{fig:FieldCascade}(a)).

\begin{figure}[h]
    \centering
    \includegraphics[width=0.7\linewidth]{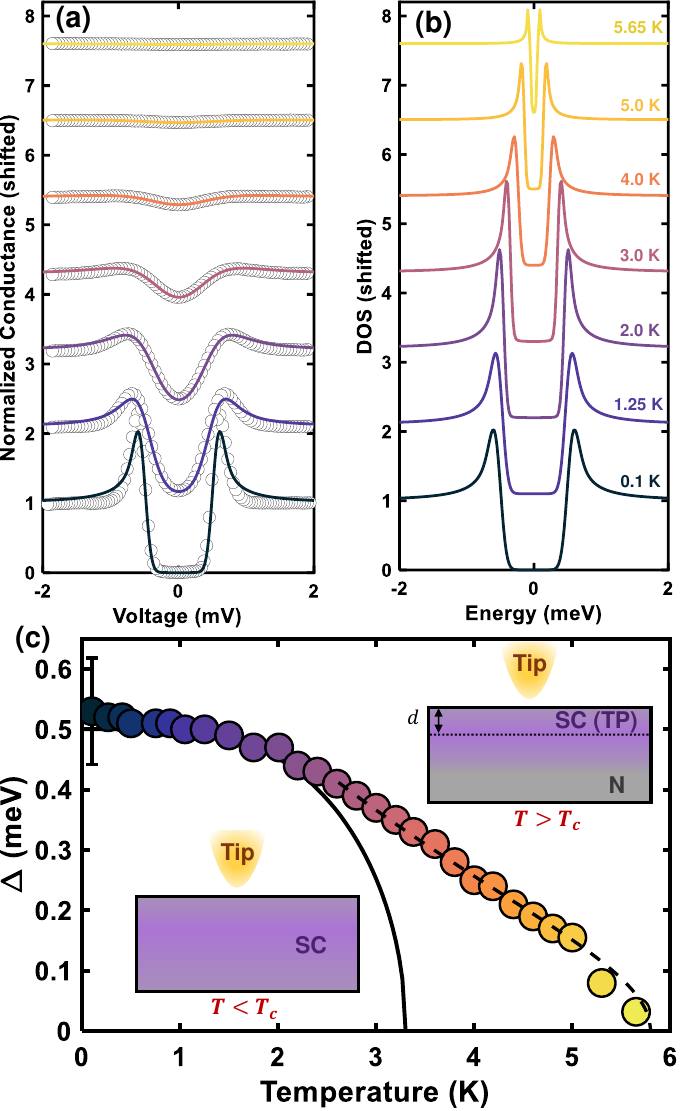}
    \caption{(a) Tunneling conductance versus bias voltage curves as a function of temperature is shown as colored lines. The black lines are the fits obtained by convoluting the density of states in (b) with the derivative of the Fermi function. The temperature is shown close to each curve. (b) Density of states obtained in a similar way as in Fig.\,\ref{fig:GapvsT1}(b). The distribution of values of the superconducting gap is here of about $0.1$ meV. (c) Superconducting gap $\Delta$ as a function of temperature. The value of $\Delta$ is determined by the gap value used to obtain the density of states in (b) as described in the text. The black line represents the expected BCS tendency for a superconducting gap $\Delta_0=0.52$ meV. The error bar provides the distribution of values of the superconducting gap used to calculate (b) and remains roughly constant with temperature. Dashed line is the expression described in the text for enhanced superconductivity close to defects. In the inset we show schematically the expected situation for a temperature above the bulk superconducting $T_{c}$ but below the $T_{c}$ close to the defect (dashed line). We assume that superconductivity (violet) is enhanced at a defect lying at a distance $d$ to the surface.We emphasize that the enhanced superconductivity may occur anywhere close to a defect, not just close to the surface.}
    \label{fig:GapvsT}
\end{figure}

In a few locations at zero magnetic field we find areas presenting  an anomalous temperature dependence of the superconducting tunneling conductance. We show the measured tunneling conductance obtained in one of such areas in Fig.~\ref{fig:GapvsT}(a) as colored lines. Again, we use the density of states shown in Fig.~\ref{fig:GapvsT}(b) to obtain the temperature dependence of the superconducting density of states with temperature, shown by thin black lines Fig.~\ref{fig:GapvsT}(a). We observe superconducting features well above the previously obtained $T_{c}\approx 3$ K, up to 5 K. The temperature dependence of the superconducting gap (points in Fig.~\ref{fig:GapvsT}(c)), obtained from the density of states shown in Fig.~\ref{fig:GapvsT}(b), does not follow BCS theory (black line in Fig.~\ref{fig:GapvsT}(c)).

\section*{Discussion}

PtPb$_4$ is a simple s-wave BCS superconductor and there are no signatures of topological features in the bulk superconducting density of states. However, our measurements did not reveal atomically flat regions on the cleaved surface. Surface states arise due to the termination of the crystal along the direction perpendicular to the surface. These states typically appear within energy gaps of the bulk dispersion and are evanescent in the out-of-plane direction, meaning that $k_z$ is no longer a good quantum number. Instead, surface states disperse along $k_x$ and $k_y$ forming a two-dimensional electron gas which can be well separated from the bulk\,\cite{ECHENIQUE1989111,RevModPhys.75.933,Herrera2023b}. Thus, rough surfaces do not allow for the establishment of a two-dimensional surface state. The formation of such surface states, as observed in PtSn$_4$\,\cite{Wu2016} requires large atomically flat terraces. Since no such regions were detected in our experiment, the establishment of a well-defined surface state is unlikely. Consequently, while we find no evidence of topological features in the superconducting state, we cannot exclude their presence in scenarios where atomically flat regions are available.

Interestingly, we find that there are regions presenting superconducting critical parameters $T_c$ and $H_{c2}$ which are considerably higher than those expected from the bulk, with $T_c$ nearly twice and $H_{c2}$ nearly five times the bulk values. AuSn$_4$ also presents an increased $T_c$ with however much smaller enhancement by about 20\%\,\cite{Herrera2023}. The significant $T_c$ enhancement we find in PtPb$_4$ is reminiscent of results in the intermetallic compound PtBi$_2$, where topological Fermi arcs are present at the surface, coinciding with superconducting features at critical temperatures several times higher than those of the bulk\,\cite{Schimmel2024,Kuibarov2024,Gao2018}. The increased values of $T_c$ we find here in PtPb$_4$ suggests that there are situations close to the surface in PtPb$_4$ with superconducting properties that could be radically different from the bulk properties.

It is relevant to note that the layered structure of PtPb$_4$ allows for the formation of polytypes with different stacking arrangements of Pb-Pt-Pb layers. The structures \#68 ($Ccce$) and \#125 (\pnbm) only differ in the stacking arrangement of Pb-Pt-Pb groups and are difficult to distinguish unambiguously using X-ray diffraction data (we provide x-ray scattering results in the Appendix)\,\cite{Lee2021}. It seems thus important to understand the electronic properties on layers with different Pb-Pt-Pb arrangements, and the role of the interface between these arrangements at defects such as stacking faults or twinning planes. The role of twinning planes in the $T_c$ was investigated in single crystals of several sizes in elemental superconductors. Having different electron-phonon coupling at a twinning plane was assumed to lead to an increase in the electron-phonon coupling constant $\lambda$ localized around the defect\,\cite{Buzdin1984,Buzdin1985,Abrikosov1988,Khlyustikov1988}. Point contact spectroscopy measurements have confirmed such an enhancement in electron-phonon coupling\,\cite{Khotkevich90}. Since this increase is primarily localized along the defect, it typically results in only a minor increase in $T_c$ and $H_{c2}$. However, in small superconducting grains, the influence of defects is maximized, potentially leading to a logarithmic increase of $T_c$\,\cite{Buzdin1984}. Estimations in Sn lead to an average electron-phonon coupling that can produce in principle a ten-fold increase in $T_c$. Observations in granular Sn and other elemental metals show indeed an enhancement by a factor of two to three\,\cite{Buzdin1984,Khlyustikov83,Khlyustikov1988}, compatible with results of an increased electron-phonon coupling\,\cite{Khotkevich90}.

Interestingly, the temperature dependence of the superconducting gap in the regions with increased $T_c$ in PtPb$_4$ does not follow expectations from BCS theory (Fig.~\ref{fig:GapvsT}(c)). We can estimate the temperature dependence of the observed gap in the tunneling conductance by using the Ginzburg-Landau solution for the temperature dependence of the superconducting order parameter in a system with enhanced twinning plane superconductivity. From Ref.\,\cite{Khlyustikov87} we find that, above the bulk critical temperature $T_{c}$, the temperature dependence of the order parameter $\varphi$ can be written as $\varphi=\frac{\sqrt{2t}}{sinh(\vert d\vert t^{1/2}+p)}$, where $t=\frac{T-T_{c}}{T_{cd}-T_{c}}$, $p=0.5ln\frac{1+t^{1/2}}{1-t^{1/2}}$, $T_{cd}$ the critical temperature close to the defect, and $d$ the distance to the defect, with $d=1$ at a distance of approximately the coherence length of the superconductor free of defects $\xi$. From $H_{c2}\approx 0.36$ T and $H_{c2}=\frac{\Phi_0}{2\pi\xi^2}$ we estimate a coherence length $\xi\approx 32$ nm \cite{Xu2021,ShenPhD2020}. We find a good fit to the temperature dependence of the gap obtained from the tunneling conductance (dashed line in Fig.~\ref{fig:GapvsT}(c)) if we take $d=1.2$, $T_{c}=2.4$ K and $T_{cd}=5.8$ K. This shows that the superconducting $T_c$ can be considerably enhanced close to structural defects. These defects can lie several tens of nm below the surface. This suggests that there are portions of the sample with different amounts of defects, separated by stacking faults and leading to the patch-like patterns found in the density of states above $H_{c2}$ (Fig.~\ref{fig:FieldCascade}).

\section*{Conclusions}

In summary, we find that PtPb$_4$ is a s-wave BCS superconductor with a negligible density of states close to the Fermi level and a superconducting gap following BCS expectations. PtPb$_4$ possibly has large networks with different amounts of defects, created by small energy difference between different stackings of Pb-Pt-Pb layers. These networks create a patch-like patterns of superconducting and normal regions close to the surface presenting $T_c$ and $H_{c2}$ that can be considerably enhanced with respect to the bulk. A-priori preparation of interfaces of PtPb$_4$, for example using deposition, and further experiments on samples of PtPb$_4$ prepared under different conditions are very interesting prospects in view of better understanding interface physics related to superconductivity\,\cite{10.1093/nsr/nwae272}. These could lead to both enhanced superconductivity and the establishment of surface states with topological properties. Furthermore, the interesting prospects raised by Rashba spin-splitting and anisotropy at the surface\,\cite{Lee2021}, might be leveraged by atomically flat surfaces.

\section*{Acknowledgments}
Some samples of PtPb$_4$ were obtained in the framework of a practical teaching activity at the Universidad Aut\'onoma de Madrid. We acknowledge the enthusiasm and participation of the 2023 year students in the Master's degree in Condensed Matter Physics and Biological Systems. We acknowledge support by the Spanish State Research Agency (PID2020-114071RB-I00, PID2023-150148OB-I00, CEX2023-001316-M, TED2021-130546B-I00), by the Comunidad de Madrid through program NMag4TIC-CM (Program No. TEC-2024/TEC-380), and by EU (VectorFieldImaging Grant Agreement 101069239 and COST superqumap CA21144). Segainvex and Sidi at UAM, Madrid, are acknowledged for help in STM construction and sample characterization. Work done at Ames Laboratory was supported by the U.S. Department of Energy, Office of Basic Energy Science, Division of Materials Sciences and Engineering. Ames Laboratory is operated for the U.S. Department of Energy by Iowa State University under Contract No. DE-AC02-07CH11358. 

\section*{Appendix}

\begin{figure}
    \centering
    \includegraphics[width=1\linewidth]{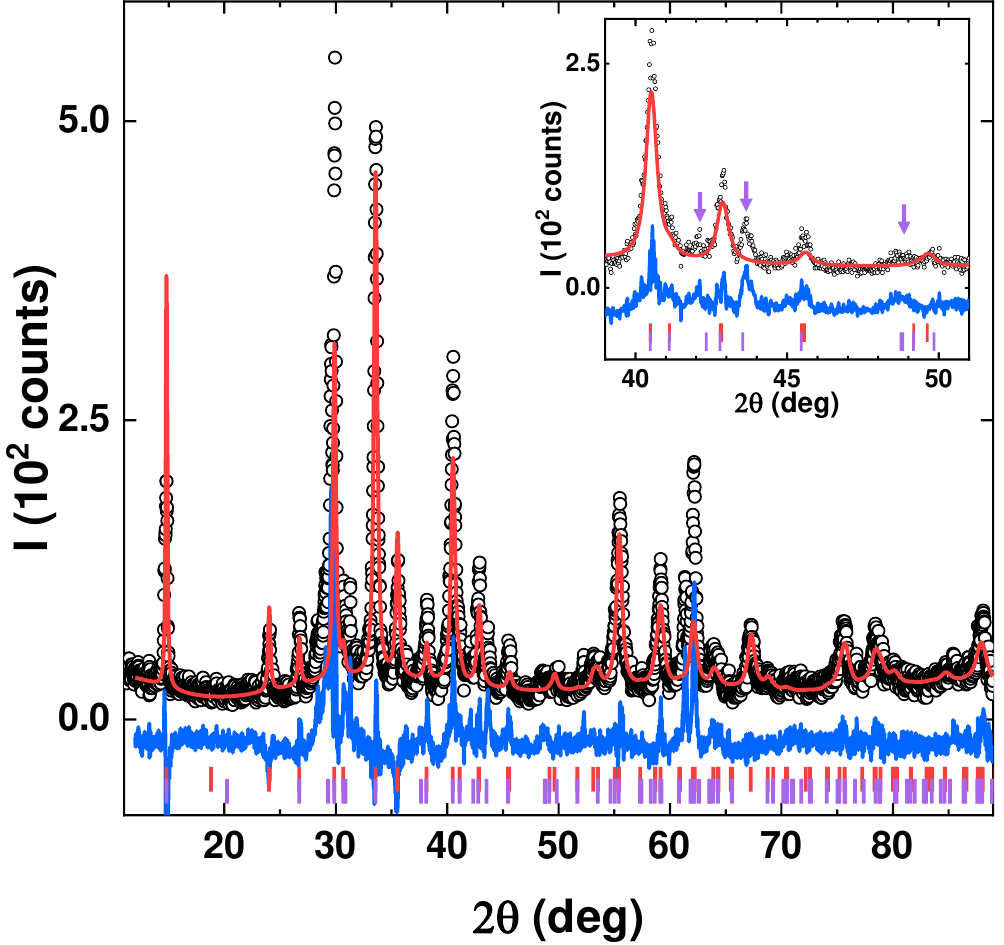}
    \caption{X-ray diffraction pattern of PtPb$_4$ powder is shown by black circles. The best pattern obtained with the \pnbm \ structure by refinement is shown by a red solid line. The difference between the experimental and the refined diffraction pattern is shown by a blue line. Main peaks of the \pnbm \ structure are shown as red vertical lines, while peaks of the \textit{Ccce} structure are shown as purple vertical lines.  Inset shows some peaks in our data that could correspond to the \textit{Ccce} structure.}
    \label{fig:RX}
\end{figure}

We note that high energy x-ray diffraction data were acquired in Ref.\,\cite{Lee2021}. However, for completeness we provide in Fig.~\ref{fig:RX} the results of the x-ray scattering data in our samples. Powder X-ray Diffraction was performed using a Bruker D8 Discover diffractometer. A small set of single crystals were finely grounded into powder and deposited evenly on the surface of a Si wafer before XRD measurements. Resulting diffraction data was analyzed with a Rietveld refinement using FULLPROF software\,\cite{fullprof}. The refinement shows a good fit to the \#125 (\pnbm) space group, yielding lattice parameters of $a=b=6.6663(9)$ \AA, and $c=5.9780(3)$ \AA. Data are also compatible with the \#68 ($Ccce$) space group, showing that these are very similar and just differ by changes in the stacking of layers of Pb-Pt-Pb\,\cite{Lee2021}.

%

\end{document}